# Forecasting of Meteorological variables using statistical methods and tools

*Emmanuel Agbo*

*Cross River University of Technology, Calabar, Cross River State Nigeria*
*Lafarge Africa PLC, Mfamosing Plant, Akamkpa, Cross River State, Nigeria*
*Department of Physics, Osun State University, Osogbo, Osun State, Nigeria*

*Corresponding Author address: Calabar, Cross River, Nigeria*
*Corresponding Author email: emmanuelpaulagbo@gmail.com*

## Abstract

The need to understand the role of statistical methods for the forecasting of climatological parameters cannot be trivialized. This study gives an in depth review on the different variations of the Mann-Kendall (M-K) trend test and how they can be applied, regression techniques (Simple and Multiple), the Angstrom-Prescott model for solar radiation, etc. The study then goes ahead to apply some of them with data obtained from the Nigerian Meteorological Agency (NiMet), and applying tools like the python programming language and Wolfram Mathematica.

Results show that the maximum ambient temperature for Calabar is increasing (Z=2.52) significantly after the calculated p-value < 0.05 (significant level). The seasonal M-K test was also applied for the dry and wet seasons and both were found to be increasing (Z=3.23 and Z=4.04 respectively) after their calculated p-values < 0.05.

The relationship between refractivity and other meteorological parameters relating to it was discerned using partial differential equations giving the gradient of each with refractivity; this was compared with results from the correlation matrix to show that the water vapour contents of the atmosphere contributes significantly to the variation of refractivity.

Multiple linear regression has also been adopted to give an accurate model for the prediction of refractivity in the region after the residual error between the calculated refractivity and predicted refractivity was minimal.

**Keywords:** Meteorology, Forecasting, Python Programming, Climate, Mann-Kendall, Multiple Linear Regression.

## 1. Introduction

The importance of statistical modelling and forecasting of time series data, etc., cannot be overemphasized. The benefits ranges from easy interpretability arising from visualization of results to the removal of the mysticism factor for the layman. The word 'forecasting' has to do with predicting the future based on data from the past and present. This is regularly done by the analysis of trends.

A routine example might be the estimation of temperature trends for some specified future date. Compared to forecasting, prediction can be seen as a term which is more general.

Forecasting methods have been applied in different areas ranging from climatology, finance, foreign exchange, etc. This has been applied in different regions of the world for better prediction and simulation. The key distinction in Information and Communication Technology (ICT) is the fact that with this technology, we can make predictions and simulations from previously obtained data. This is true and can be applied for every area while paying attention to the rules that govern them.

In this study we will be applying some statistical methods which can be adopted for the forecasting of climatic (weather) parameters in different regions of the world.

It is important to note the predictability of the atmosphere is not perfect, this brings into context the fact that although statistical methods are necessary, results obtained are not totally accurate which is why room for errors (uncertainties) are given, albeit, a trend can be observed [1]. Statistical methods have been applied in the study of different regions for example, Daniel S. Wilks in [1] buttressed on the use of these methods on the analyses of different regions that do not necessarily have the same climatic conditions. This brings into context the fact that laws are true irrespective of the region, i.e. neglecting all other factors that have little contribution to weather, the same methods can be applied in different regions to yield accurate results.

Analysis of trends can be useful in depicting and predicting the changing patterns and erraticism of some climatic parameters. This analysis gives a proper knowledge about the changing conditions of the climate and its effects, by the evaluation of meteorological parameters.

A data scientist using any tool or software for modelling and forecasting is particular interested with the progression of these parameters (meteorological) as a function of time(t) $f(t)$. The designers of navigation or monitoring systems cannot trivialize the importance of forecasting as this is a very important part of their system. The spatial and temporal changes of atmospheric parameters call for the adoption of this analysis to discern the effects of some meteorological parameters on some variables; for example, see [2].

A very popular software for any data scientist that is willing to understand the nitty-gritty of weather forecasting is Python Programming. This paper will explain in detail the setup processes for this to help the layman get started. A dataset of temperature trend in Calabar, Nigeria will be used at the end of this chapter to test the processes explained for better visualization.

The applicability of results from forecasting cannot be underestimated because this is great information for people that depend on weather conditions like farmers, surfers, and event planners, etc. The accurate prediction of atmospheric parameters can go a long way in positively affecting the financials of the informed as money can be saved by avoiding unnecessary cost during trying times [3]. Natural disasters like Tsunami can be predicted with the correlation of meteorological parameters, harnessing information as explained previously and the incorporating this information through machine learning into the design of forecasting systems.

We delve deeper into a review of statistical methods like the M-K test and its different variations, the Angstrom-Prescott model for the estimation of solar radiation, linear regression techniques, with a deep look into multiple linear regression which will be applied in the predicting refractivity after obtaining the coefficients of the variables. Results will be obtained and explained.

## 2. Review of Statistical Tests/Methodology

With the shift going on in the world of technology, the implementation of some time series forecasting will be explained as well as the python implementation technique. We often use forecasting models on time series data for the estimation of future trends of meteorological parameters.

### 2.1 Statistical Test for Trend (Mann-Kendall trend test)

One of the most important and widely applied test for trends involving time series is the Mann-Kendall trend test. It is mostly used for environmental and hydrological data. The test is non parametric and does not necessitate the data conforming to a particular distribution, similarly, the sensitivity of the test due to an inhomogeneous series resulting to abrupt breaks is very low [4]. The null hypothesis $H_o$ which says that there is no monotonic trend in the series, is tested against the alternative hypothesis $H_1$ which says that there is a trend in the series. The test is applied to cases where a range of data $x_i$ is in agreement with the equation below;

$$x_i = f(t_i) + \varepsilon_i \tag{1}$$

$f(t_i)$ is a function of time and $\varepsilon_i$ are the range residuals with zero mean.

The Mann-Kendall test statistic $S$ is calculated using the formula

$$S = \sum_{k=1}^{n-1} \sum_{j=k+1}^{n} \text{sgn}(x_j - x_k) \tag{2}$$

where;

$$\text{sgn}(x_j - x_k) = \begin{cases} +1; & \text{if } (x_j - x_k) > 0 \\ 0; & \text{if } (x_j - x_k) = 0 \\ -1; & \text{if } (x_j - x_k) < 0 \end{cases} \tag{3}$$

$n$ in eq. (2) is the number of data values in the studied series. The advantage of this test is that it can handle the situation where data a values are incomplete with respect to the number of years or months, etc. [4]

In the case where $n$ is greater than or equal to 10 (10 and above), we adopt the normal approximation ($Z$).
To find the variance of $S$, '$VAR(S)$', we compute eq. (4) below.

$$VAR(S) = \frac{1}{18}\left[ n(n-1)(2n+5) - \sum_{p=1}^{g} t_p(t_p-1)(2t_p+5) \right] \tag{4}$$

From the equation, the number of data values is represented by $n$, the number of equal of tied groups is represented by $g$, and the number of data values in the $p^{th}$ group is represented by $t_p$.
We now use the results from $VAR(S)$ to find the test statistic $Z$

$$Z = \begin{cases} \dfrac{S-1}{\sqrt{VAR(S)}}; & S > 0 \\ 0; & S = 0 \\ \dfrac{S+1}{\sqrt{VAR(S)}}; & S > 0 \end{cases} \quad (5)$$

A decreasing trend can be discerned from results of eq. (5) when the value of $Z$ is negative and an increasing trend when $Z$ is positive.

| Significance level ($\propto$) | Required n |
|---|---|
| 0.1 (10%) | ≥ 4 |
| 0.05 (5%) | ≥ 5 |
| 0.01 (1%) | ≥ 6 |
| 0.001 (10%) | ≥ 7 |

**Table 1.** Significance level ($\propto$) required for given numbers of data

The significance of an increasing or decreasing trend is observed when the p-value of the series is lower than the significance level ($\propto$)**,** in this case, we can say that there is an trend observed trend in the series. [5]

The classification of this probability/significance level is important because results can be confused to be entirely true. We need to understand that the significance level of say 0.05, means that there is a 5% probability that a mistake will be made while rejecting the null hypothesis $H_o$. Similarly, a significance level of 0.01 means that there is a 1% probability that a mistake will be made while rejecting $H_o$

## 2.2 Regression analysis

The two easiest ways to forecast time series data by observation are the simple regression and the moving average, they both depend on historical data. The former demands mere observation of the previous trend and drawing up an extrapolation from there; this can be somewhat less accurate. The moving average has been used for forecasting meteorological data like rainfall (See reference [6]). Analyzing with regression has to do with the relationship one variable which is dependent has with one or more independent variables. We use them to check for models showing the strength of relationship between the variables and any possible future relationships [1].

### 2.2.1 Simple Linear Regression
This regression variation is based on the assumption that the two variables (dependent and independent variable) show a linear relationship between the intercept and the slope, similarly, there is no residual error in this regression and the value is constant across all observations.

$$Y = \pm mX \pm c + e \quad (6)$$

Y is the dependent variable
X is the independent variable
m is the value of the slope
c is the intercept
x e is the residual error
The regression is depicted by a straight line describing the eq. (6) above.

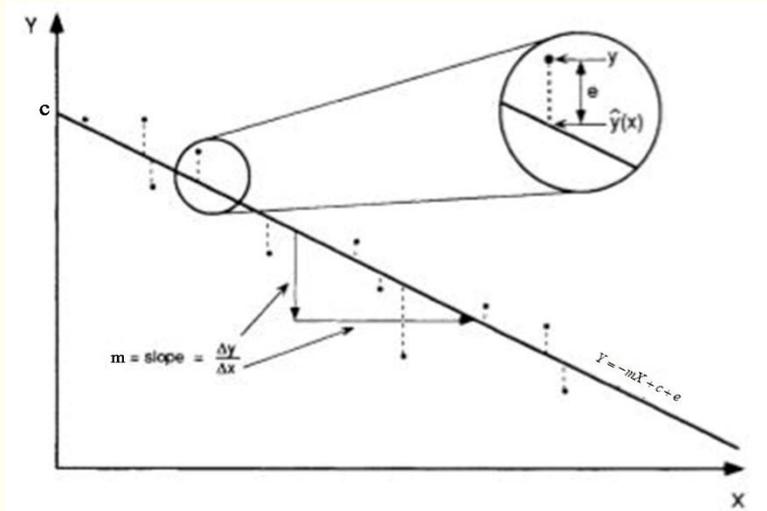

**Figure 1.** *Schematic illustration of simple linear regression. The regression line, $Y = -mX + c + e$, is chosen as the one minimizing some measure of the vertical differences (the residuals) between the points and the line. The residual e is the difference between the data point and the regression line.*

### 2.2.2 Multiple Linear Regression

This model is similar to that of simple linear regression, but the only exception is that it has multiple independent variables, unlike that of simple linear regression which has just the one. This can be represented by eq. (7);

$$Y = \pm m_1 X_1 \pm m_2 X_2 \pm m_3 X_3 \pm c + e \tag{7}$$

Y is the dependent variable
$X_1, X_2, X_3$ are the independent variables
$m_1, m_2, m_3$ are the values of the slopes
c is the intercept
e is the residual error

One thing to note about multiple linear regression is that the independent variables must not be collinear, i.e., they don't have to have a high correlation coefficient between each other, else there will be difficulty in assessing the relationship between the dependent and independent variables.

We also need to take note that before multiple linear regression is performed on range of data values, a linear relationship must exist between the independent and dependent variables. The amount of residual error must be almost constant at each point in the model. The multiple linear regression will be applied to study and predict refractivity trend in Calabar, Nigeria. This was done with the 'statsmodel' package in python programming and results have been displayed in section (2.5)

A perfect meteorological equation that this regression technique can be applied to is the refractivity equation recommended by the International Telecommunication Union (ITU) shown in eq. (8);

$$N = 77.6\frac{P}{T} + 3.73 \times 10^5 \frac{e}{T^2} (N-units) \tag{8}$$

*P* is the Atmospheric Pressure (hPa)
*e* is the Atmospheric Vapour Pressure (hPa)
*T* is the Absolute Temperature (K)

Eq. (8) shows the relationship between refractivity (dependent variable) and meteorological parameters (ambient temperature, atmospheric pressure, and vapour pressure) which are all independent variables.

This has been applied in [7] modelling the meteorological parameters for the accurate determination of refractivity. These meteorological parameters (Ambient Temperature, Atmospheric Pressure and Relative Humidity) has been obtained from the Nigeria meteorological Agency (NiMet), Calabar.

Results have been presented in section (2.5). From eq. (8), we obtain the atmospheric vapour pressure *e* from the relation;

$$e = \frac{e_s H}{100} (hPa) \tag{9}$$

$e_s$ is the saturated vapour pressure (hPa) calculated from;

$$e_s = 6.11 \exp\left(\frac{17.26(T-273.16)}{T-35.87}\right)(hPa) \tag{10}$$

## 2.3 Review of the Application of Simple Linear Regression Analysis in Climatology (The Angstrom-Prescott model)

The linear regression technique can be applied to find the relationships between an independent variable and the dependent variable. We can see the explanation of this from eq. (6).

One major example of the benefits of linear regression is the estimation of the Angstrom-Prescott coefficients of the Angstrom-Prescott model for a particular region as this relates to solar radiation. The Angstrom-Prescott model is given by [8];

$$\frac{H}{H_0} = a + b\frac{n}{N} \tag{11}$$

where the monthly average daily extraterrestrial radiation is given by $H_0$, $H$ is the monthly average daily global radiation in Wh/m$^2$/day. $n$ is the actual sunshine duration in a day for a particular region (hours), $N$ is the monthly mean length of the day in hours. The Angstrom-Prescott empirical coefficients are given by *a* and *b*. The linear regression technique has been adopted by Srivastava and Pandey [8] to find by *a* and *b*. Comparing eq. (6) to eq. (9) we have that;

$$\frac{H}{H_0} = Y \text{ (variable)}$$

$$\frac{n}{N} = X \text{ (variable)} \tag{12}$$

$$b = m = slope$$

$$a = c = Y \text{ intercept}$$

This shows that if we have the variables '$\frac{H}{H_0}$ and $\frac{n}{N}$', we can get the values of $a$ and $b$, from our Y intercept and slope respectively. Getting these constant values for specific regions will help us forecast future trends.

For better understanding, the extraterrestrial radiation $H_0$ is given by the equation [9];

$$H_0 = \frac{24 \times 3600 \times I_{SC}}{\pi} \times \left[1 + 0.33\cos\left(\frac{360 \times d}{365}\right)\right] \times \left[\cos\phi\cos\delta\sin\omega + \frac{\pi\omega}{180}\sin\phi\sin\delta\right] \quad (13)$$

Here, $I_{SC}$ is the solar constant with a value of 1367W/m², $d$ represents the day of the year (from January 1st to December 31st); taking January 1st as 1 and December 31st as 365 or 366 (in the case of a leap year). The latitude of the study location, the declination angle and the sunset hour angle are represented by $\phi, \delta$, and $\omega$ respectively. $\omega = \cos^{-1}(-\tan\phi\tan\delta)$. The declination angle can be obtained from [9]

$$\delta = 23.45\sin\left[360\left(\frac{284+d}{365}\right)\right] \quad (14)$$

The monthly mean length of the day (in hours) can be obtained from [9]

$$N = \frac{2\omega}{15} \quad (15)$$

The above equations can be applied to estimate the coefficients using linear regression. By this we can use these coefficients to predict solar radiation for a given region.

We know that the declination angle ranges from $-23.5 \le \delta \le +23.5$. From figure 3, we can see that the declination angle is $0°C$ at the Verbal and Autumnal Equinox, while the angles are -23.5 and +23.5 at the summer and winter solstice respectively. It is easy to see why this has a huge effect on the variation of Global solar radiation.

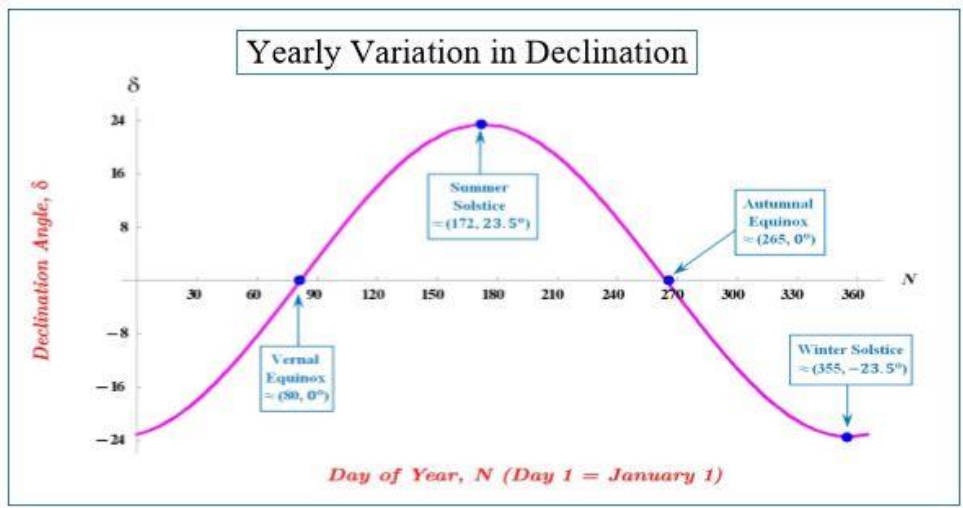

Figure 2. Yearly variation of declination angle $\delta$ with respect to the days of the year.

Klein in 1977 [10] recommended average days of the various months and corresponding angle of declination as in Table 1.

| Month | Date | Day of the year (d) | declination angle ($\delta$) |
|---|---|---|---|
| January | 17 | 17 | -20.9 |
| February | 16 | 47 | -13 |
| March | 16 | 75 | -2.4 |
| April | 15 | 105 | 9.4 |
| May | 15 | 135 | 18.8 |
| June | 11 | 162 | 23.1 |
| July | 17 | 198 | 21.2 |
| August | 16 | 228 | 13.5 |
| September | 15 | 258 | 2.2 |
| October | 15 | 288 | 9.6 |
| November | 14 | 318 | -18.9 |
| December | 10 | 344 | -23 |

Table 2: Recommended average days for various months and their corresponding declination angles [10].

## 2.4 Calculus in Climatology

Applying calculus in environmental science is important in predicting a lot of things. It can be applied to understand the impacts of parameters on the variations of other parameters that they relate to. It is important to know that calculus is the 'mathematical study continuous change' so this can be applied in climatology to discern the impacts of some parameters on the "continuous change" of others [11-13]

Writing the refractivity equation in terms of relative humidity $H$, by substituting (10) into (9), and the into (8), we have;

$$N = 77.6\frac{P}{T} + 3.73 \times 10^5 \frac{6.11\exp\left(\frac{17.26(T-273.16)}{T-35.87}\right) \times 0.01H}{T^2} \quad (N-units) \qquad (16)$$

Similarly, obtaining refractivity in terms of the saturated vapour pressure $e_s$ using eq. (8) and (9) gives;

$$N = 77.6\frac{P}{T} + 3.73 \times 10^5 \frac{e_s H}{100T^2} (N-units) \qquad (17)$$

Now applying partial differentials to the equations for refractivity; eqs. (8), (16), and (17), we obtain partial differentials relating each parameter to refractivity;

$$\frac{\partial N}{\partial P} = \frac{77.6}{T}$$

$$\frac{\partial N}{\partial T} = -\left(77.6\frac{P}{T^2} + 7.46\times 10^5 \frac{e}{T^3}\right)$$

$$\frac{\partial N}{\partial H} = \frac{22790.3\exp\left[\dfrac{17.26(-273.16+T)}{-35.87+T}\right]}{T^2} \qquad (18)$$

$$\frac{\partial N}{\partial e} = \frac{3.73\times 10^5}{T^2}$$

$$\frac{\partial N}{\partial e_s} = \frac{3.73\times 10^3 \times H}{T^2}$$

From monthly Temperature, Humidity and Atmospheric pressure data obtained for 2005-2018 from the archives of the Nigerian meteorological agency (NiMet) Calabar, the atmospheric vapour pressure and the saturated vapour pressure can be obtained by applying these parameters in eqs. (9) and (10).

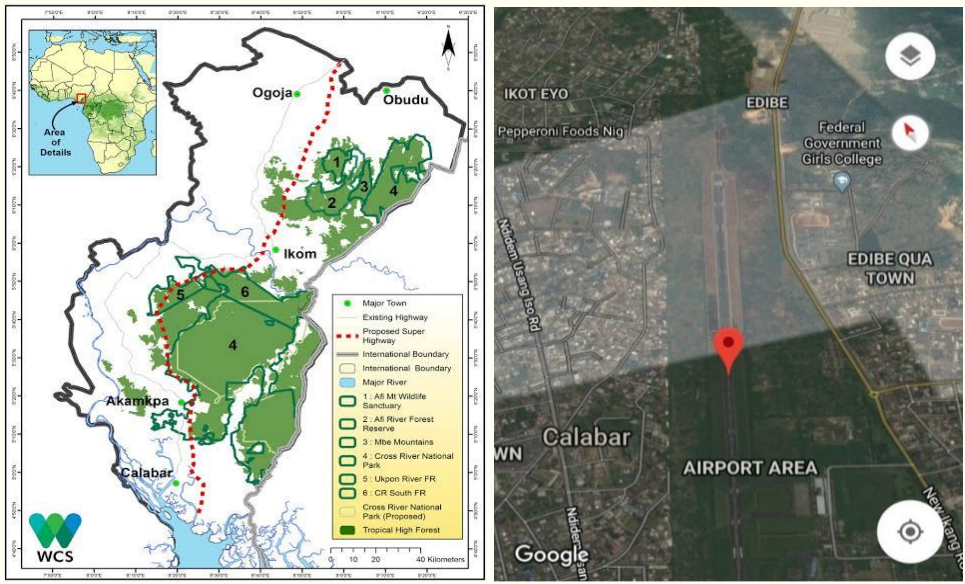

Figure 3. Map of study area showing Calabar as a coastal area (left) and the exact location of the Nigerian Meteorological Agency (NiMet) where the data was obtained (right).

## 2.5 Python implementation for Mann-Kendall trend test

With the python software installed, the next step will be installing an IDE (integrated development environment). The easiest IDE to use is the Jupyter Notebook. This IDE displays results as you code.

We will walk you through the processes for analyzing data by using the data for Calabar in the south of Nigeria, collected from the archives of the Nigeria meteorological agency (NiMet). Research has been done in this area in climatology [14-18], but with the application of python and the Mann-Kendall test can give more meaning to time series data.

We need to install the python package for the Mann-Kendall test called '*pymannkendall*'. To install this package, the following python packages are required;
- Numpy
- Scipy

For handling and cleaning data we need the '*pandas*' package, and for data visualization we need the '*matp*lotlib' package.

We want to analyze maximum ambient temperature data for 20 years in Calabar.
In the Jupyter notebook, the first step will be to import the respective packages. We must also note that for our examples in the Appendices, we stored the excel file containing the data used for the analysis in the same folder as the python file for easy reference.

*Appendix A* shows the process of importing the installed packages required for the analysis into the workspace.

Before we perform the Mann-Kendall test, we need to import the excel file titled '*Temperature'* in which the table is stored, in a sheet name called '*MAX*'. See *Appendix B*

*Appendix C* shows how the Mann-Kendall original test is performed after importing the packages and data. We assigned the name of the imported data file as '*Max*' and set the significance level (∝) to the default 5% (0.05); this can be adjusted by the user to his/her preference. Results were obtained and tabulated.

We now perform the seasonal M-K test for the dry season variation, we import the excel file titled '*Temperature',* the date column will be an index column. The sheet name of the excel file in which the data is stored is called 'dry'. This implementation can be seen from *Appendix D*

*Appendix E* shows the seasonal M-K test python implementation for the dry season variation. By setting the significance level (∝) to the default 5% (0.05), and the period to 4, which stands for the 4 months of the dry season in the study area (November to February), we have satisfied the criteria for the seasonal M-K test.

For the wet season variation, the excel file titled '*Temperature'* will be imported and the date will be an index column. The sheet name is called 'wet'. *Appendix F* shows the implementation code for this importation.

We can now perform the seasonal Mann-Kendall test on the wet season data. *Appendix G* shows this. The Seasonal Mann-Kendall test of the imported file we assigned the name 'wet' has been achieved by setting the significance level (∝) to the default 5% (0.05); this can be adjusted by the user to his preference. We also set the period to 8, which stands for the 8 months of the wet season in the study area (March to October).

There are other variations of the Mann-Kendall test along with their python implementation [19]. These can be used depending on the data obtained and the aim of the test.

1. Hamed and Rao Modified MK Test (*hamed_rao_modification_test*): This test addresses serial correlation issues

2. Yue and Wang Modified MK Test (*yue_wang_modification_test*): This is also a variance correction method for considered serial autocorrelation proposed by Yue, S., & Wang, C. Y. (2004). User can also set their desired significant n lags for the calculation.

3. Modified MK test using Pre-Whitening method (*pre_whitening_modification_test*): This test pre-whitens the time series before applying the trend test

4. Modified MK test using Trend Free Pre-Whitening method (*trend_free_pre_whitening_modification_test*): This test removes the trend component from the series before pre-whitening and the applying the trend test

5. Multivariate MK Test (*multivariate_test*): As the name implies, this test is for multivariate (multiple) parameters. This can be used for monthly data, where each month can be considered as a parameter.

6. Regional MK Test (*regional_test*): As the name implies, this calculates the trend at a regional scale

7. Correlated Multivariate MK Test (*correlated_multivariate_test*): Unlike the Multivariate MK test, this test is also a multivariate mk test, but the parameters are correlated.

8. Correlated Seasonal MK Test (*correlated_seasonal_test*): This test is similar to the seasonal MK test, but in this is used when the time series is significantly correlated with previous seasons/months

9. Partial MK Test (*partial_test*): Due to the fact that in some studies, many factors can affect the dependent parameters, so we overcome this by inputting one dependent parameter and an independent parameter.

10. Theil-Sen's Slope Estimator (*sens_slope*): This test method proposed by Theil (1950) and Sen (1968) [20] is applied to estimate the magnitude of the monotonic trend.

11. Seasonal Theil-Sen's Slope Estimator (*seasonal_sens_slope*): This test method considers the seasonal effect of the Theil-Sen's Slope Estimator.

## 3. Results and Discussion

### 3.1 Results

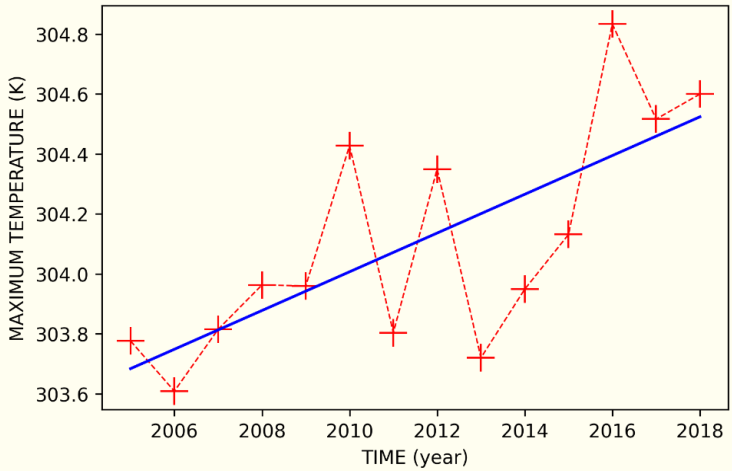

Figure 4. Mann-Kendall trend of Maximum Ambient Temperature

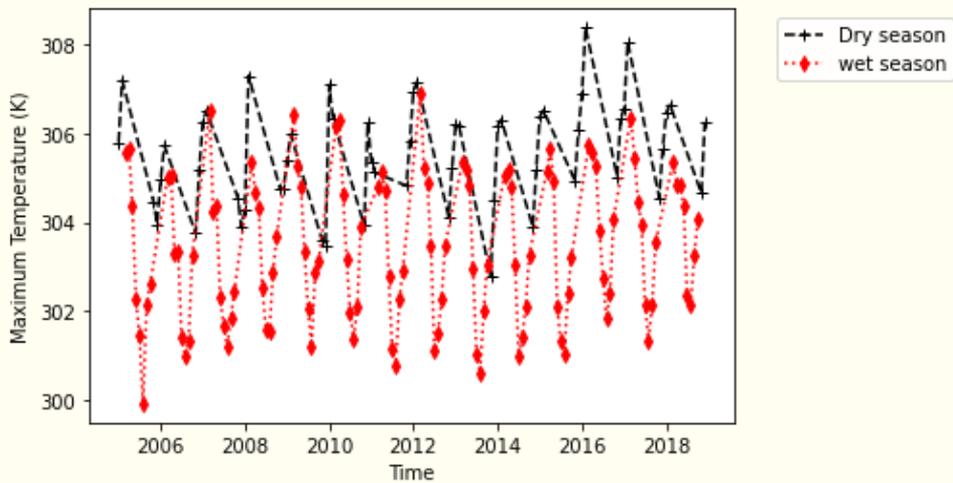

Figure 5. Seasonal trend of Maximum Ambient Temperature for dry and wet season.

### 3.2 Discussion

For the annual variation in figure 4, results show that there is a trend in the series as the p-value is less than the significance level (0.05). The positive Z value (observed from *Appendix C*) shows that the series is increasing. We can conclude that the maximum ambient temperature variation is increasing, and it is doing so with significance, the slope of the trend can be observed from the results in *Appendix C*.

For the dry season variation observed in figure 5, results show that there is a trend in the series. The positive Z value of the dry season trend observed from *Appendix E* shows that the series is increasing. We can conclude that the maximum temperature variation in the dry season is increasing significantly as the calculated p-value is greater than the significance level (0.05), the slope of the trend can be observed from results in *Appendix E*.

For the wet season variation observed also in figure 5, results show that there is a trend in the series. The positive Z value from *Appendix G* shows that the series is increasing. We can conclude that the maximum temperature variation in the dry season is increasing significantly as the calculated p-value is greater than the significance level (0.05), the slope of the trend can be observed from the results in *Appendix G*. These results are in agreement with Agbo et al. [2] for the same region.

### 3.2.1 Relationship between Refractivity and Meteorological Parameters

To understand the relationship between all refractivity and all parameters relating to it, we adopt eq. (18) by substituting obtained and calculated data

From the data obtained at the Nigerian Meteorological Agency (NiMet) Calabar, and adopting eq. (9) and (10) we obtain the total annual values for the meteorological parameters as;

$P$ = 1005.97hPa; $H$ = 85.71%; $T$ = 300.28K; $e$ = 30.71hPa; $e_s$ = 35.94hPa. Substituting these values into the equations in eq. (18), we obtain;

$$\frac{\partial N}{\partial P} = 0.258425$$
$$\frac{\partial N}{\partial T} = -0.0196183$$
$$\frac{\partial N}{\partial H} = 1.48436 \qquad (19)$$
$$\frac{\partial N}{\partial e} = 4.13672$$
$$\frac{\partial N}{\partial e_s} = 3.62832$$

Results from the gradients of the differential equations in eq. (18) show that the vapour pressure and saturated vapour pressure contributes more to the variation of refractivity. The relative humidity similarly has a high gradient; this can be physically explained by relating the water vapour content of the atmosphere to the variation of refractivity.

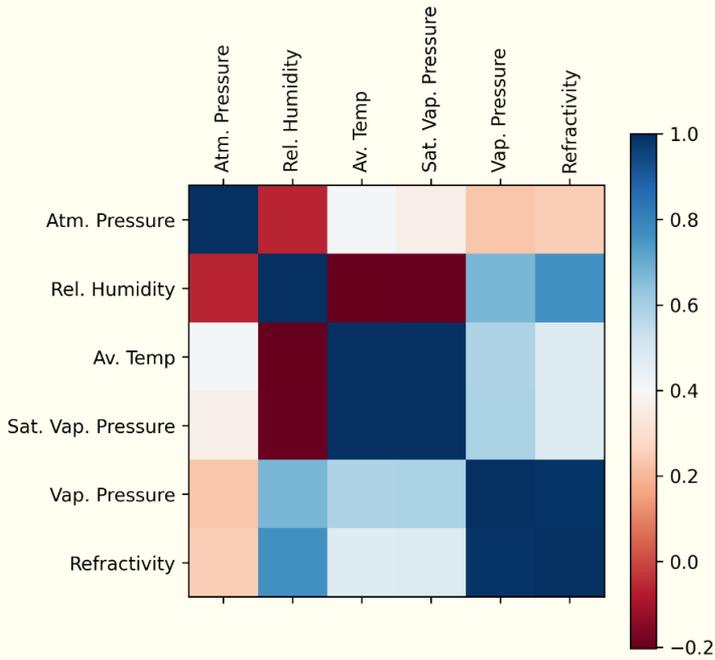

Figure 6. Correlation Matrix of atmospheric parameters and refractivity.

The correlation plot of refractivity and all other meteorological parameters is shown in figure 6. Results agree with that of the differential equations in eq. (19). As seen in eq. (19), the correlation plot showed that the atmospheric vapour pressure and relative humidity had high positive relationships with refractivity. The saturated vapour pressure however has a low correlation coefficient compared to the high gradient in eq. (19); this can be interpreted thus; that the variation of the saturated vapour pressure has a relatively high contribution to the variation of refractivity, but the saturated vapour pressure does not have a similar trend to that of refractivity.

### 3.2.2 Application of Multiple Linear Regression in Climatology

| Year | Pressure | Temperature | Humidity | Refractivity (N) |
|---|---|---|---|---|
| 2005 | 1005.15 | 300.23 | 87.15 | 388.88 |
| 2006 | 1005.38 | 300.17 | 85.38 | 385.73 |
| 2007 | 1005.50 | 300.10 | 84.84 | 384.74 |
| 2008 | 1005.44 | 300.21 | 86.00 | 387.11 |
| 2009 | 1005.83 | 300.29 | 83.36 | 383.75 |
| 2010 | 1005.46 | 300.71 | 83.26 | 385.62 |
| 2011 | 1005.80 | 300.12 | 87.49 | 388.83 |
| 2012 | 1005.75 | 300.44 | 87.97 | 391.57 |
| 2013 | 1005.74 | 300.03 | 87.07 | 387.69 |
| 2014 | 1005.92 | 299.79 | 85.15 | 383.38 |
| 2015 | 1006.55 | 300.03 | 85.68 | 385.83 |
| 2016 | 1006.97 | 300.70 | 85.69 | 389.73 |
| 2017 | 1007.09 | 300.58 | 86.33 | 389.90 |
| 2018 | 1007.02 | 300.52 | 84.58 | 386.84 |

Table 3: Data of Obtained Meteorological Parameters and Refractivity

Multiple linear regression has been applied to relate refractivity with obtained meteorological parameters. The goal is to obtain an equation that relates refractivity to meteorological parameters through Multiple Linear Regression. Using eq. (8) to calculate refractivity, we show results in table 3. As part of the conditions for carrying out multiple linear regression, we have to test for collinearity between the independent variables. We see from the correlation matrix in Figure 6 that the independent variables are not collinear, hence this satisfies the criteria for carrying out MLG.

|  | C | $S_e$ | t Stat | P-value | Lower 95% | Upper 95% |
|---|---|---|---|---|---|---|
| **Intercept** | 1617.97 | 51.05 | -31.69 | $2.30 \times 10^{-11}$ | -1731.72 | -1504.23 |
| **Pressure** | 0.17 | 0.05 | 3.25 | $8.75 \times 10^{-03}$ | 0.05 | 0.28 |
| **Temperature** | 5.68 | 0.13 | 44.90 | $7.22 \times 10^{-13}$ | 5.39 | 5.96 |
| **Humidity** | 1.53 | 0.02 | 68.62 | $1.05 \times 10^{-14}$ | 1.48 | 1.58 |

Table 4: Output of the Multiple Linear Regression showing the Coefficients (C) of each parameter and their standard error ($S_e$)

From our analysis we obtain the coefficients (slopes) of the variables (meteorological parameters) and the intercept from Table 4 to form the equation below;

$$Refractivity(N) = 1.53\,Humidity + 0.17\,Pressure + 5.68\,Temperature - 1617.97 \quad (20)$$

The above equation can be used to accurately predict the variation of refractivity, given the values of the meteorological parameters. Table (4) shows these results obtained from the multiple linear regression. The values for the predicted refractivity (N) gotten from eq. (20) but substituting the values of the meteorological parameters. This equation is more straight forward that the equation recommended by ITU as all the variables and coefficients are all linear with respect to refractivity.

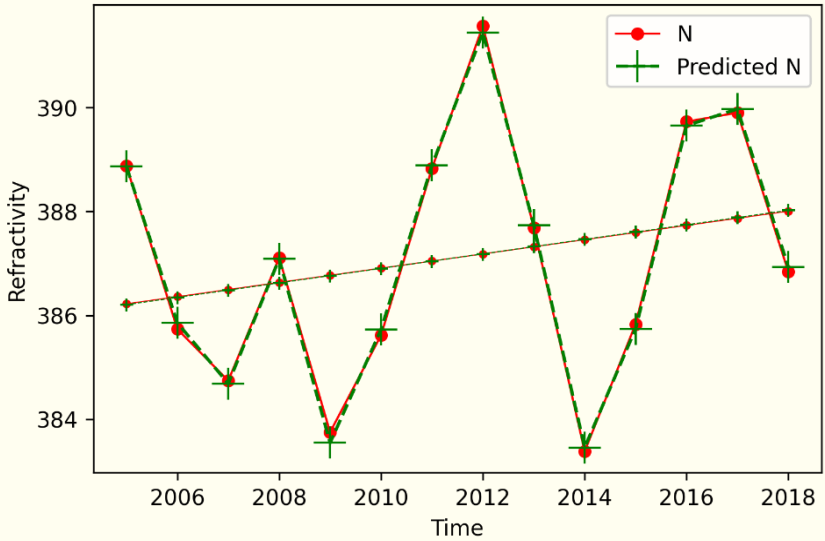

Figure 7: Comparison plot of annual refractivity and predicted refractivity.

Figure (7) shows the trend of refractivity calculated from eq. (8) with that of predicted refractivity, calculated from eq. (20). The residual error seen from table 5 shows relatively constant values (in agreement with our MLG conditions), and a small deviation from the original values of refractivity.

From table 4 probability values (p-values) of the parameters are all less than the significance level (5% = 0.05; 95% confidence level), this shows that the variation agrees with the alternative hypothesis and shows a trend relating the independent variables to the dependent variables.

| YEAR | N | Predicted N | Residuals |
|---|---|---|---|
| 2005 | 388.88 | 388.87 | 0.005 |
| 2006 | 385.73 | 385.85 | -0.121 |
| 2007 | 384.74 | 384.69 | 0.056 |
| 2008 | 387.11 | 387.09 | 0.025 |
| 2009 | 383.75 | 383.54 | 0.204 |
| 2010 | 385.62 | 385.72 | -0.109 |
| 2011 | 388.83 | 388.89 | -0.060 |
| 2012 | 391.57 | 391.45 | 0.124 |
| 2013 | 387.69 | 387.74 | -0.048 |
| 2014 | 383.38 | 383.45 | -0.074 |
| 2015 | 385.83 | 385.74 | 0.091 |
| 2016 | 389.73 | 389.65 | 0.076 |
| 2017 | 389.90 | 389.97 | -0.072 |
| 2018 | 386.84 | 386.93 | -0.095 |

Table 5: Residual Output derived from the results of the coefficients, showing the predicted refractivity values compared to the refractivity values to give the residuals.

Results from figure 7 show the minimal error between the predicted refractivity and the calculated refractivity. Table 5 shows the values for both as well as the residual error between them. This shows that the error is small and thus, eq. (20) can be adopted for the prediction of refractivity for the study area. This equation can be modified and refractivity N can be gotten in terms of other parameters like the saturated vapour pressure and the atmospheric vapour pressure.

**4.0 Conclusion**

There are myriads of ways in which weather can be forecasted and this arises from the understanding of basic meteorological parameters and how they behave in the atmosphere; and also from the understanding of the role of statistics in climate research [21]. Research in this area has been reviewed to give a better understanding of the different techniques for analyzing trends; which include, Linear Regression (Multiple and Simple), the Mann-Kendall trend test [22, 23] (to test for trends in a time series variation), the Angstrom-Prescott model for estimating solar radiation as well as the python implementation of some various techniques.

The multiple linear regression technique was applied to model an equation to accurately predict the trend for refractivity in the study location, the simple linear regression technique has been explained as well as accurate methods for its application in the predicting/estimation of the Angstrom-Prescott coefficients. These coefficients can be gotten for specific regions and can be accurately applied to predict solar radiation in that region.

Results from the multiple linear regression gave an accurate model for the prediction of refractivity in the region after the residual error between the calculated refractivity and predicted refractivity was minimal.

The Mann-Kendall original and seasonal test has been applied to analyze the maximum temperature in Calabar, Nigeria for the annual and seasonal (dry and wet season) variation respectively, and results show that the annual, dry season and wet season had increasing variations (after having positive Kendall Z-values of 2.52, 3.23, 4.04 respectively) and they were all increasing significantly at 5% (0.05) level of significance after their p-values were all less than 0.05 agreeing with Agbo and Ekpo [23].

The relationship between refractivity and other meteorological parameters relating to it was discerned using partial differential equations giving the gradient of each with refractivity; this was compared with results from the correlation matrix to show that the water vapour contents of the atmosphere contributes significantly to the variation of refractivity.

**Acknowledgments**

The author will like to acknowledge the Nigerian Meteorological Agency (NiMet) Calabar for providing the necessary data for applying in this study.
The author will also like to express his thanks and appreciation to the editor, whose comments greatly improved the chapter.

## Conflict of Interest

The author declares no conflict of interest

## Appendices

### Appendix A
**Input:**
```
import numpy as np
import pandas as pd
import pymannkendall as mk
import matplotlib.pyplot as plt
%matplotlib inline
from pandas import ExcelWriter
from pandas import ExcelFile
from matplotlib.figure import Figure
```

This will import the above installed packages into the workspace.

### Appendix B
```
Max = pd.read_excel("Temperature.xlsx", 'MAX'
index_col= 'YEAR')
```

The excel file titled '*Temperature*' will be imported and the data will be an index column. The sheet name is called '*MAX*'.
We can now perform the Mann-Kendall test

### Appendix C
**Input**
```
mk.original_test(Max, alpha=0.05)
```

**Output**
```
Mann_Kendall_Test(trend='increasing', h=True,
p=0.011793457077065028, z=2.518264946676251,
Tau=0.5164835164835165, s=47.0,
var_s=333.6666666666667, slope=0.06763844012453053,
intercept=303.5218288324476)
```

### Appendix D
```
dry=pd.read_excel("Temperature data.xlsx", 'Sheet2',
index_col= 'YEAR')
```

The excel file titled '*Temperature*' will be imported and the data will be an index column. The sheet name is called 'dry'.
We can now perform the Mann-Kendall test

### Appendix E
**Input**
```
mk.seasonal_test(dry, alpha=0.05, period=4)
```
**Output**
```
Seasonal_Mann_Kendall_Test(trend='increasing',
h=True, p=0.001232892414896325, z=3.231159219618304,
Tau=0.3269230769230769, s=119.0,
var_s=1333.6666666666667, slope=0.08467049808428379,
intercept=305.1036046113848)
```

#### Appendix F

```
wet=pd.read_excel("Temperature data.xlsx", 'Sheet3',
index_col= 'YEAR')
```

The excel file titled '*Temperature'* will be imported and the data will be an index column. The sheet name is called 'wet'.

We can now perform the Mann-Kendall test

#### Appendix G

**Input**
```
mk.seasonal_test(wet, alpha=0.05, period=8)
```
**Output**
```
Seasonal_Mann_Kendall_Test(trend='increasing',
h=True, p=5.1261530983781161e-05,
z=4.049799512953561, Tau=0.28846153846153844,
s=210.0, var_s=2663.3333333333335,
slope=0.05741935483871145,
intercept=302.85004032258064)
```